\documentclass[a4paper,12pt]{article} 
\usepackage{graphicx} 
\usepackage{graphics} 
\usepackage{amsmath} 
\usepackage{amsfonts}  %necessaire aux caracteres spéciaux
\usepackage{amssymb} %necessaire aux caracteres spéciaux

\usepackage[colorlinks=true, pdfstartview=FitV, linkcolor=blue,  citecolor=blue, urlcolor=blue]{hyperref}  %liens hypertexte

\newcommand{\Pl}{\mathbb{P}} %Planck
\def\l{\left}
\def\r{\right}
\def\be{\begin{equation}}
\def\ee{\end{equation}}
\def\beq{\begin{equation}}
\def\eeq{\end{equation}}

\begin{document}
%%%%%%%%%%%

%titres possibles: On the neutrino and electron masses
\title{On the neutrino and electron masses in the theory of scale relativity}
\author{Laurent Nottale\\{\small LUTH, CNRS, Paris Observatory and Paris University} \\{\small 92195 Meudon CEDEX, France}\\
{\small laurent.nottale@obspm.fr}}
\maketitle

\begin{abstract}
%%%%%%%%
We have long ago derived a theoretical relation between the mass of the electron and the fine structure constant \cite{Nottale1994}, which writes to lowest order $\alpha \ln  ({m_\Pl}/{m_e})  =3/8$  (where $m_\Pl$ is the Planck mass). We suggest the existence of a similar relation valid for neutrinos, $\alpha  \ln  ({m_\Pl}/{m_\nu} ) =1/2$. 
 From this relation, we theoretically predict a lightest neutrino mass $m_{\nu}=m_\Pl \: \exp (  -\alpha^{-1}/2 )=0.0214$ eV. The masses of the two heavier neutrinos, $0.0231$ eV and $0.0552$ eV, can then be obtained from  experimental results of neutrino oscillations.
 \end{abstract}
 %%%%%%%

%%%%%%%%%%%
\section{Introduction}
%%%%%%%%%%

The discovery of neutrino oscillations (see  Ref.~\cite{PDG2020} and references therein) proved that neutrinos are not massless. However, only differences of square masses are known through these experiments and observations, not the masses themselves.

The current view about the origin of elementary particle masses attributes them to their Yukawa coupling to the Higgs field. This is based on the fact that the coupling constants and the masses are equal. However, one may remark that the acquisition of masses by the $W_\pm$ and $Z_0$ bosons through the spontaneous breaking of the $SU(2) \times U(1)$ electroweak (EW) symmetry leads to an explicit prediction of these masses, while this is not the case for the fermion masses. One can therefore interpret these facts in the reverse way, namely, that the fermion masses are determined by a different mechanism and that they subsequently define the Yukawa couplings with the Higgs field.

Among the reasons which lead one to suspect that the HIggs / spontaneous symmetry breaking mechanism is not sufficient to understand fermion masses, there is the huge difference between the $e,\,\mu,\, \tau$ masses and that of their associated neutrinos, despite their similar electroweak quantum numbers. Actually, the neutrinos should be precisely massless with the Standard Model \cite{PDG2020}. On the other hand, it is remarkable that the masses of $\nu_e$, $e$ and $u,\,d$ quarks are in increasing order while they respectively are subjected to the (weak), (weak and electromagnetic), then (weak, electromagnetic and strong) forces. This suggests that the origin of fermion masses could be linked to the gauge fields themselves.

In the theory of scale relativity \cite{Nottale1992,Nottale1993,Nottale2011}, a geometric interpretation of the nature of gauge transformations, of gauge fields and their associated charges has been proposed \cite{Nottale1994,Nottale2006}, \cite[Chapt.~7]{Nottale2011}. In this framework, one naturally obtains an universal relation between couplings (i.e. square of charges) and mass scales ratios with the Planck scale (or equivalently Compton-length-scale ratios). 

Such a relation between the electron to Planck mass ratio and the fine structure constant has been long ago proposed \cite{Nottale1994}. In the present paper, we suggest the existence of a similar relation for neutrinos, which allows us to obtain a theoretical prediction for the lightest neutrino mass, then to derive the other masses from experimental results of neutrino oscillations.

\section{Mass-charge relation for the electron}
%%%%%%%%%%%%%%%%%%%%%%%

\subsection{Gauge theories in scale relativity}
%*********
In the theory of scale-relativity and fractal space-time \cite{Nottale1992,Nottale1993,Nottale2011}, the geometry of space-time is generalized to a non-differentiable continuum. This implies its fractality, i.e. its explicit dependance on the resolution scale (going as far as divergence when the scale interval tends to 0). In this framework, we add the space-time resolutions (which may become realized through the measurement resolutions) to the fundamental variables that characterize in a relative way the state of the coordinate system (in addition to position, orientation and motion) \cite{Nottale1989,Nottale1992,Nottale1993}. Then particles are identified with the geodesics of this fractal space-time: more specifically, the velocity field ${\cal V}$ of these geodesics provides a geometric interpretation of the nature of the wave function  (e.g. $m{\cal V}= - i \hbar \nabla \ln \psi$ in the simplest case where $\psi$ is solution of a Schr\"odinger equation)  \cite{Nottale1993,Nottale2006,Nottale2007,Nottale2011}.

This has led us to a new interpretation of gauge transformations and gauge fields \cite{Nottale1994, Nottale2006,Nottale2011} in which the variable conjugate to the charge is just the {\it resolution scale}. In other words, charges find their origin, according to Noether's theorem, in the symmetries of the space of resolution-scales (internal to the geodesics which are identified to the particle).

As a consequence the wave function of a free electron can be written as:
\beq
\psi= \psi_0 \times e^{i[(px-Et+\sigma \phi )/\hbar+ 2 \pi \, \widetilde{\alpha}(r) \ln ({\lambda_e}/r) ]},
\eeq
where $\widetilde{\alpha}(r)$ is a running coupling, $r$ is a running resolution length-scale and $\lambda_e=\hbar/m_e c$ is the Compton scale of the electron. We know that angles $\phi$ vary between 0 and $2 \pi$, so that spin differences  $ \delta \sigma$ are universally quantized in units of $\hbar$. In the same way, in the theory of special scale-relativity (SSR) where the Planck length-scale $\lambda_\Pl$ is re-interpreted as a minimal, undepassable scale of resolution, invariant under dilations and contractions \cite{Nottale1992}, the scale variable is limited as $\ln (\lambda_e/r) \leq \ln ({\lambda_e}/{\lambda_\Pl})$, so that the charge is quantized according to the relation:
\beq
\widetilde{\alpha} \; \ln \frac{\lambda_e}{\lambda_\Pl}=1.
\label{mcr1}
\eeq
We have suggested, since the running occurs up to the Planck scale, that the coupling appearing in this relation should be the effective electromagnetic coupling of the electroweak theory, $\tilde{\alpha}=8 \alpha/3$, where $\alpha$ is the fine structure constant. This mass-coupling relation can be understood as yielding the mass of the electron from its charge.

\subsection{Bare charge $1/2\pi$}
%***********

The electron charge itself can be derived through its running from its infinite energy value (i.e. Planck length-scale value in SSR) to the electron energy $0.511$ MeV. In the Standard Model (SM), the value of the charge at infinite energy is divergent. This is no longer the case in SSR, where an infinite energy corresponds to the finite Planck length-scale (in similarity with special motion relativity, where an infinite value of energy corresponds to a finite value $v=c$ of velocity), according to the log-Lorentz relation \cite{Nottale1992,Nottale1993}:
\beq
\ln \frac{m}{m_e}=\frac{\ln ({\lambda_e}/{\lambda})}{\sqrt{1-\ln^2 ({\lambda_e}/{\lambda})/\ln^2 ({\lambda_e}/{\lambda_\Pl})}},
\eeq
where the referernce scale has be taken here to be the electron scale.
It is easy to check in this formula that $m \to \infty$ when $\lambda \to \lambda_\Pl$ and that the Planck length $ \lambda_\Pl=\sqrt{\hbar G/c^3}$ becomes, in the SSR framework, a limit scale, unreachable and invariant under dialtions.

We have argued \cite{Nottale1996,Nottale1993,Nottale2011} that the infinite energy value (which can be identified with  its ``bare" value) of the inverse electromagnetic coupling (including the electroweak factor 3/8) is expected to have the value $4 \pi^2$. Let us briefly recall the argument.

The force between two charges can be computed as:
\begin{equation}
F =\frac{\delta p} {\delta t},
\end{equation}
where $\delta p$ is the momentum exchanged by the intermediate boson during the interaction time $\delta t$. Since the boson is assumed to be of zero mass and moving at light velocity, the time interval is $\delta t={r}/{c}$,
where $r$ is the distance between the two charges. 

This force can be decomposed into a radial and an anguler part, $F=F_r \times F_{\theta\phi}$.
As concerns the radial part of the force, it can be established by describing the exchanged boson in terms of an harmonic oscillator, according to the second quantization theory. A general method making use of the concept of information entropy has been devised by Finkel \cite{Finkel1987} for constructing any exact Heisenberg relation between any couple of variables and to determine the kind of wave function which achieves the limit of the inequality. 

In the situation considered here, $\delta x=r$ is an {\it interval} of distance. In this case, the Finkel method \cite{Finkel1987} allows one to show that the Heisenberg relation writes $r\: \delta p \geq \hbar/\pi$, and that the limit $\hbar/\pi$ is just reached by the harmonic oscillator distribution. 
Therefore we find that the radial part of the momentum variation in the exchange of null mass bosons over a distance $r$ is given by
\beq
\delta p= \frac{\hbar}{\pi r}.
\eeq
As regards the angular part of the force, it is naturally given by spherical harmonics $|Y_l^m(\theta,\phi)|^2$, i.e. in the isotropic case considered here, by
\beq
|Y_0^0(\theta,\phi)|^2=\frac{1}{4\pi}.
\eeq
The force created by the exchange of null mass bosons writes
\beq
F=\frac{\hbar c} {\pi r^2} \times \frac{1}{4\pi}=\frac{\hbar c} {4\pi^2 r^2}.
\label{5}
\eeq
By comparing Eq.~(\ref{5}) with the standard expression of the force written in terms of a coupling constant, $F =  {\alpha \, \hbar c}/ {r^2}$, one finally obtains a ``natural" value for the coupling constant \cite{Nottale1996}:
\beq
\alpha_\infty^{-1}=4 \pi^2= 39.478...
\eeq
We have shown \cite{Nottale1993,Nottale1996,Nottale2011} that the running of the effective electromagnetic coupling in the electroweak theory, $\alpha_0(r)^{-1}=8 \alpha(r)^{-1}/3= \frac{5}{3}\alpha_1(r)^{-1}+\alpha_2(r)^{-1}$ fairly supports such an expectation: we obtained $\alpha_{0\infty}^{-1}=39.489 \pm 0.013$ in \cite[Sec.~11.1.3.4]{Nottale2011}, which differs from $4 \pi^2$ by less than 1$\sigma$.

\begin{figure}[!ht]
\begin{center}
\includegraphics[width=12cm]{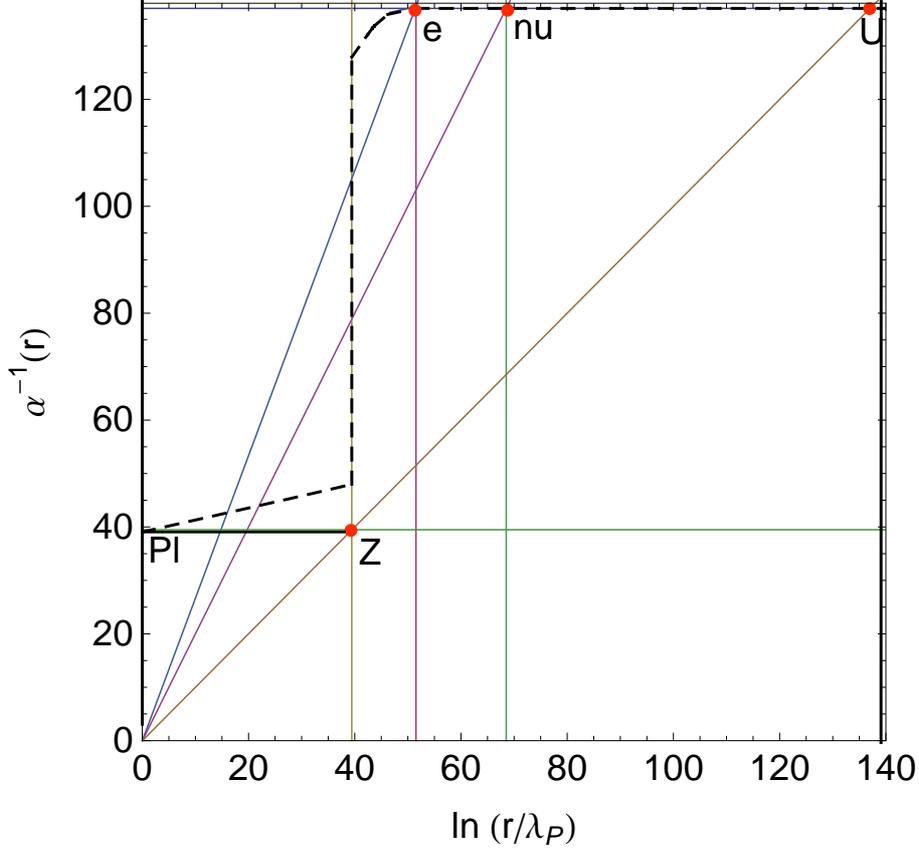}%from MASS-CHARGE-RELATION.nb
\caption{\small{
Scale-coupling relations in a global diagram ranging from the Planck length-scale to the Universe length-scale (defined by the cosmological constant \cite{Nottale2019}). In the SSR theory, the Planck  length-scale is a minimal scale, invariant under dilatations. The natural value of an inverse coupling at this scale is $4 \pi^2$ (see text, horizontal black line). The electromagnetic inverse coupling starts from this value ($39.478$), runs toward the $WZ$ electroweak scale where it jumps by a factor $8/3$ and reaches the value $\approx 128$ (because of the Higgs mechanism leading to the emergence of the $W_+,W_-$ and $Z_0$ masses), and then increases due to the effects of the various elementary particle pairs, up to the electron scale where it acquires its large scale value $137.036$ (black dashed thick curve). Four scale-coupling relations, which can be written as $\ln ({r}/{\lambda_\Pl})=\widetilde{\alpha}^{-1}$, are considered in this diagram. (1) Taking $\widetilde{\alpha}^{-1}=4 \pi^2$ yields the $WZ$ scale (marked Z in the diagram). (2) With $\widetilde{\alpha}^{-1}= \alpha^{-1}$, one obtains a cosmological scale (U in the diagram). (3) With $\widetilde{\alpha}^{-1}=(3/8) \alpha^{-1}$, where $\alpha$ is the low energy fine structure constant,one gets the electron scale. (4) Finally, taking $\widetilde{\alpha}^{-1}=(1/2) \alpha^{-1}$, one obtains a new length-scale that can be identified with the lightest neutrino Compton-length (nu in the diagram), yielding a predicted mass of $0.0214$ eV.
}}
\label{run}
\end{center}
\end{figure}

The running of this coupling from this value ($4 \pi^2$) at infinite energy scales (i.e. Planck length-scale in SSR) to low energy scales determines the low energy value of the electric charge, i.e. of the fine structure constant (this running depends on the whole content of charged elementary particles). Then its intersection with the running mass fixes the mass of the electron to its low energy value (see Fig.~\ref{run}).

Note that this universal value $4 \pi^2$ of an inverse coupling at infinite energy can also be combined with the mass-coupling relation Eq.~(\ref{mcr1}) to yield a fundamental mass-scale \cite{Nottale1996,Nottale2011} given by
\beq
\ln\frac{m_\Pl}{m_{WZ}}=4 \pi^2.
\eeq
This mass-scale is $1.397 \times 10^{17}$ smaller than the Planck mass, i.e. $m_{WZ}=87.4$ GeV \cite{Nottale1996}, which is typical of the electroweak scale since it is intermediate between the $W$ ($80.379$ GeV) and $Z$ boson ($91.1876$ GeV) masses. Such a relation may therefore help solving the hierarchy problem (i.e. the large ratio between the GUT scale, which is the Planck mass-scale in SSR, and the electroweak scale).

Another application of this mass-coupling relation concerns the strong inverse coupling, which can be shown, in the SSR framework, to reach the value $4 \pi^2$ just at the scale where it crosses the gravitational inverse coupling \cite{Nottale2011}.

\subsection{Lowest order mass-coupling relation}
%***********
The ratio of Compton lengths between electron and Planck scales can be replaced by the mass ratio:
\beq
\ln \frac{\lambda_e}{\lambda_\Pl}=\ln \frac{m_\Pl}{m_e}.
\eeq
Therefore, the mass-charge relation we theoretically obtained for the electron \cite{Nottale1994} reads:
\beq
\frac{8}{3} \, \alpha \, \ln  \frac{m_\Pl}{m_e}  =1.
\eeq
The present recommended values of the fine structure constant and of the electron mass \cite{CODATA2018,PDG2020} are
\beq
\alpha^{-1}=137.035999084(21), \;\;\; m_e=0.51099895000(15) \times 10^{-3}\; {\rm GeV},
\eeq
while the Planck mass determination has been now improved to \cite{CODATA2018,PDG2020} 
\beq
m_\Pl=1.220910(29) \times 10^{19}\; {\rm GeV}.
\eeq
These values yield numerically
\beq
\frac{8}{3} \, \alpha \, \ln  \frac{m_\Pl}{m_e}  =1.0027,
\eeq
which differs by less than $0.3\:\%$ from the predicted value $1$.

\subsection{Account of threshold effects on the running charge}
\begin{figure}[!ht]
\begin{center}
\includegraphics[width=10cm]{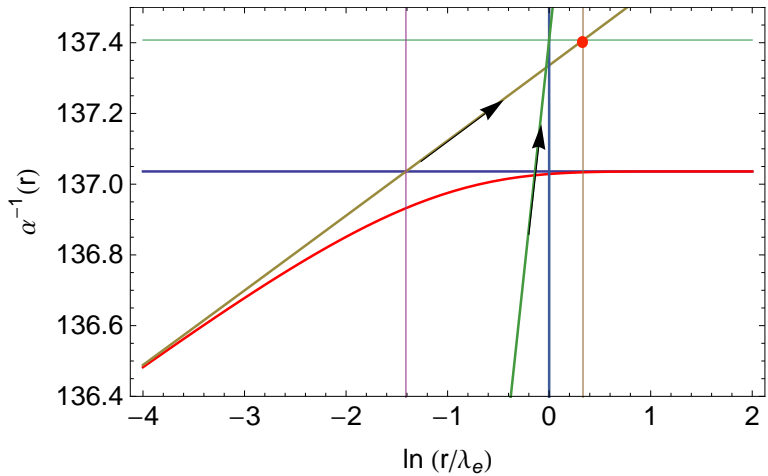}%MASS-CHARGE-RELATION.nb
\caption{\small{
Behavior of the electron scale-coupling relation around the electron Compton scale, including the threshold effect (drawn here in terms of length-scale instead of mass-scale). Red curve: running electromagnetic inverse coupling during the electron-scale transition toward the inverse fine structure constant $\alpha^{-1}=137.036$. Green line: running scale factor $ (8/3) \ln(r/\lambda_\Pl$, which reaches the value of the inverse electromagnetic coupling at the electron scale. Brown line: asymptotic inverse running coupling $\alpha_{as}^{-1}(r)= \alpha^{-1} + \frac{2 }{3 \pi}   \l(\ln(r/\lambda_e)+\gamma+\frac{5}{6}\r) ]$; it crosses $\frac{8}{3} \ln\frac{m_\Pl}{m_e}=137.408$ (green horizontal line) at the scale where the running electron mass is assumed to reach its low energy constant value, $m(r)=m_e$ (red point, $\ln(r/\lambda_e)=-1/3$).
}}
\label{alphaas}
\end{center}
\end{figure}

%************
However, this zeroth order calculation does not account for the threshold effects that occur at the electron scale. The fine structure constant is a large length-scale value measured at the atom scales. Actually, the electron mass and the electron charge both increase below the Compton length of the electron $\lambda_e$ toward small length-scales (i.e., beyond the electron mass $m_e$  toward large mass-scales) because of radiative corrections. Therefore, there is a smooth transition around the electron Compton-length between the constant large scale value and the variable small scale running coupling, which is known since Uehling and Serber first calculation in 1935 to diverge asymptotically only logarithmically (see Fig.~\ref{alphaas}). Consequently, the scale at which the asymptotic running coupling reaches the value of the fine structure constant $\alpha$ differs from $\lambda_e$.

The mass-charge relation is actually an equation relating the running electron coupling and electron mass in function of the scale $r$
\beq
\frac{8}{3} \, \alpha(r) \, \ln  \frac{m_\Pl}{m(r)}  = 1,
\label{mcr}
\eeq
whose solution is expected to be the electron scale. Therefore, the quantities entering in this relation are the asymptotic functions, which involve offsets, given by numerical constants, with respect to the precise electron Compton scale.
The asymptotic running QED coupling is given by \cite{Landau4,Itzykson1980}
\beq
\alpha_{as}(r)=\alpha \l\{  1+ \frac{2 \alpha}{3 \pi}   \l(\ln \frac{\lambda_e}{r} -(\gamma+\frac{5}{6})\r)\r\},
\eeq
where $\gamma=0.5772156...$ is Euler's constant. This expression means that the asymptotic running coupling reaches the value of the fine structure constant at a mass scale $\exp(\gamma+5/6)=4.098\approx 4$ times the electron mass. This result is quite in agreement with the fact that the scale dependance of the coupling comes from the $e^+ e^-$ pairs of mass $2 m_e$, which therefore define the center of the transition between the two regimes (logarithmically scale-dependant at high energies vs constant at low energies).
Therefore 
 \beq
\alpha_{as}(r=\lambda_e)=\alpha \l\{  1- \frac{2 \alpha}{3 \pi}   \l(\gamma+\frac{5}{6}\r)\r\}.
\eeq
Using this value in the mass-charge relation yields:
\beq
\frac{8}{3} \, \alpha \l\{  1- \frac{2 \alpha}{3 \pi}   \l(\gamma+\frac{5}{6}\r) \r\} \, \ln \frac{m_\Pl}{m_e}  =1.00052,
\eeq
an improvement by a factor $\approx 5$ from $2.7 \times 10^{-3}$ to $5.2 \times 10^{-4}$ of the agreement between the theoretical prediction and the experimental value. 

%1%%%%%%%
%%%%%%%%

\subsection{Account of threshold effects on the running mass}
%**********
There is also a numerical constant in the expression for the logarithmic divergence of the running mass \cite{Weisskopf1939,Itzykson1980},
%vérifier si la constante est chez Weisskopf
\beq
m_r(m)=m_e \l\{1+ \frac{3 \alpha}{2 \pi} \l( \ln \frac{m}{m_e}+\frac{1}{4}  \r)  \r\}.
\eeq
The value of $\alpha_{as}$ should therefore be taken, no longer at exactly the electron Compton length $\lambda_e=\hbar/m_e c$ itself, but at the scale where $m=m_e$, i.e. $m=m_e \times e^{-1/4}=0.7788 \, m_e$, corresponding to a length-scale $\ln \lambda=\ln \lambda_e + 1/4$. One obtains a corrected relation:
\beq
\frac{8}{3} \, \alpha \l\{  1- \frac{2 \alpha}{3 \pi}   \l(\gamma+\frac{5}{6}+\frac{1}{4}\r) \r\} \, \ln \frac{m_\Pl}{m_e}  =1.00013,
\eeq
still improved by a factor $\approx 4$ and now reaching the $10^{-4}$ precision.

However, with the now improved value of the Planck mass, which is the main uncertainty in this calculation (coming in its turn from the badly known Newtonian constant of gravitation), this result remains far from error bars. One gets $\delta \ln (m_\Pl/m_e)=\delta m_\Pl/m_\Pl=2.4 \times 10^{-5}$, so that with $8 \alpha/3\approx 1/50$, the experimental current uncertainty is $4.6 \times 10^{-7}$, a value which can be expected to be still improved in the near future.

\subsection{Possible improvement of the running mass contribution}
%%%%%%%%%%%%
An improvement of this relation can be obtained by realizing that the effective running mass intervening in the full Lagrangian has a slightly different constant \cite{Itzykson1980} from the $1/4$ present in the mere mass renormalization:
%Itzykson Zuber p. 335, cf ELECTRON-NEUTRINO-MASS.nb
\beq
m_r(m)=m_e \l\{1+ \frac{3 \alpha}{2 \pi} \l( \ln \frac{m}{m_e}+\frac{3}{8}  \r)  \r\}.
\eeq
By taking again the value of $\alpha_{as}$ at the scale where $m_r(m)=m_e$, one obtains a still improved relation:
\beq
\frac{8}{3} \, \alpha \l\{  1- \frac{2 \alpha}{3 \pi}   \l(\gamma+\frac{5}{6}+\frac{3}{8}\r) \r\} \, \ln \frac{m_\Pl}{m_e}  = 0.999939,
\eeq
which differs by only $6 \times 10^{-5}$ from the expected value 1, thus achieving a new improvement by a factor $\approx 2$ of the agreement with the theoretical prediction.

From this relation one can derive a theoretical expectation for the mass of the electron from the fine structure constant:
\beq
m_e(pred)=m_\Pl \; \exp \l[  -\frac{3}{8}  \:\alpha^{-1}  \l(1+\frac{2 \alpha}{3 \pi} \l(\gamma+\frac{29}{24} \r) \r)  \r]=0.5096\, {\rm MeV},
\eeq
which lies within $0.3 \%$ of the experimental value $m_e=0.5110$ MeV.

\subsection{Next order}
%*********

A new improvement could be expected by going to the next order in $\alpha$ expansion. By taking the running EM coupling to two loops \cite{deRaphael1974} and assuming that the correct constant determining the scale at which $m=m_e$ is $1/3$ (intermediate between 1/4 and 3/8), we would obtain (see Fig.~\ref{alphaas}):
\beq
\frac{8}{3} \, \alpha \l[  1-\l( \frac{2 \alpha}{3 \pi}+\frac{\alpha^2}{2 \pi^2} \r)   \l(\gamma+\frac{7}{6}\r) \r] \, \ln \frac{m_\Pl}{m_e}  -1=-7 \times 10^{-7},
\eeq
which is of the order of size of the main uncertainty coming from the Planck mass 
($\frac{8}{3} \alpha \; \delta  m_\Pl  / m_\Pl \approx 5 \times 10^{-7}$). 

This opens the hope that we actually deal with an exact relation and that we just need to get a correct and precise description of the electron Compton transition between constant mass and charge at large length-scales to their scale dependance toward small scales.

\section{Generalization to neutrino masses}
%%%%%%%%%%%%%%

As recalled hereabove, the mass-coupling relation obtained for the electron reads to lowest order (i.e. disregarding small corrections due to threshold effects):
\beq
\frac{8}{3} \; \alpha \; \ln  \frac{m_\Pl}{m_e}  =1,
\label{mcr}
\eeq
where $\alpha=\alpha_e$ is the low energy fine structure constant.

%2%%%%%%%
%%%%%%%%

The factor $8/3$ comes from the electroweak theory. It results from the fact that the photon $\gamma$, which carries the electromagnetic interaction, is a low energy residual of the four high energy electroweak bosons from the $SU(2)_L$ field ($W_1, \, W_2, \, W_3$) and $U(1)_Y$ field ($B$). At $WZ$ scale ($\approx 90$ GeV), these bosons are combined through the Higgs mechanism into three massive bosons ($W_\pm$ and $Z_0$),  this spontaneous symmetry breaking leaving only the $U(1)_{EM}$ field, carried by the massless photon $\gamma$. 

As a consequence the electromagnetic coupling, which can be defined as a linear combination of the $U(1)$ ($\alpha_1$) and $SU(2)$ ($\alpha_2$) couplings,
\beq
\alpha_0^{-1}=\frac{5}{8} \: \alpha_1^{-1}+\frac{3}{8} \: \alpha_2^{-1},
\eeq
is abruptly decreased by a factor $3/8$ at the $WZ$ scale, leaving the running fine structure ``constant"
\beq
\alpha^{-1}=\frac{8}{3}\: \alpha_0^{-1}=\frac{5}{3} \: \alpha_1^{-1}+\alpha_2^{-1}.
\eeq
The running electron mass-charge relation Eq.~(\ref{mcr}) is therefore just expressed in terms of this high energy coupling $\alpha_0(r)$, i.e. it can be written as
\beq
 \ln  \frac{m_\Pl}{m(r)}  =\frac{5}{8} \: \alpha_1(r)^{-1}+\frac{3}{8} \: \alpha_2(r)^{-1},
 \label{newform}
\eeq
where $r$ is a running length-scale. Extrapolated to low energies, it is solved in terms of the electron mass and the electron charge \cite{Nottale1994,Nottale2011}.

Under its form Eq.~(\ref{mcr}), this kind of mass-charge relation seems not to be applicable to the neutrino, which is devoid of electric charge. However, it owns $SU(2)$ and $U(1)_Y$ charges similar to that of the electron (and constitutes a doublet with the left-handed electron) so that the EW form Eq.~(\ref{newform}) can also be meaningful for the neutrino. Namely, $t_3=-1/2$ and $y=-1$ for the left-handed electron, so that $Q_e=t_3+y/2=-1$ while $t_3=1/2$ and $y=-1$ for the neutrino, so that $Q_\nu=0$.

%This factor also corresponds to the ratio of squared charges of all fermions (quarks and leptons) over leptons, $[3 \times 3 \times((1/3)^2+(2/3)^2))+3]/3=8/3$, for 3 families of quarks in 3 colors and 2 charges $1/3$ and $2/3$ and 3 leptons in 2 charges, 1 and 0.

The electron, having a weak and electric charge, interacts with the 4 bosons. The neutrino, on the contrary, being electrically neutral, interacts with only the 3 weak bosons, $W_\pm$ and $Z_0$, but not the $\gamma$. We therefore suggest that a multiplicative factor $3/4$ should be applied to the coupling in the above relation, i.e a factor $4/3$ to the inverse coupling, in order to apply it to the lowest mass neutrino: 
\beq
 \ln  \frac{m_\Pl}{m}  =\frac{5}{6} \: \alpha_1(m)^{-1}+\frac{1}{2} \: \alpha_2(m)^{-1},
\eeq
where $m$ is a running mass. Extrapolating to low energies, this expression can be written as $ \frac{5}{6}\alpha_1(m)^{-1}+\frac{1}{2}\alpha_2(m)^{-1}=\frac{1}{2} \alpha(m)^{-1}$, which becomes constant below the electron mass-scale where we recover the standard fine structure constant $\alpha$. This finally results in the following conjectured mass-coupling relation for the neutrino:
\beq
2 \; \alpha \, \ln  \frac{m_\Pl}{m_\nu}  =1.
\eeq

This relation allows us to derive a lowest neutrino mass from the fine structure constant and the Planck mass:
%ELECTRON-NEUTRINO-MASS.nb
\beq
m_{\nu1}=m_\Pl \: \exp \l(  - \frac{1}{2 \alpha} \r)=1.7499 \times 10^{-30} \:m_\Pl=0.0214 \:{\rm eV}.
\eeq
The two other neutrino masses can now be derived from neutrino oscillation experimental results. From $\Delta m_{21}^2=(7.39 \pm 0.20) \times 10^{-5}$ eV$^2$ and 
$\Delta m_{32}^2=(2.51 \pm 0.03) \times 10^{-3}$ eV$^2$  \cite{PDG2020}, one obtains
%vérifier les valeurs 2020; OK; barres d'erreur ? difficile à estimer: dépend à nouveau des effets de seuil etc.;
\beq
m_{\nu2}=0.0231 \:{\rm eV}, \;\;\; m_{\nu3}=0.0552 \:{\rm eV},
\eeq
i.e. a mass ordering $m_{\nu1} \approx m_{\nu2} < m_{\nu3}$ and a total mass of $0.0997\approx 0.1$ eV. 

These values are quite compatible with our current knowledge about neutrinos (in particular, getting for one of the neutrino masses  $\approx 0.05$ eV occurs in several scenarios) and in good agreement with a Majorana origin of the mass which requires $0.016<\Sigma_j m_{\nu j}<(0.061-0.165)$ eV \cite{PDG2020}.

\section{Conclusion}
%%%%%%%%%%

In this paper, we have recalled the mass-coupling relation we derived for the electron in the scale-relativity framework \cite{Nottale1994}, $\alpha \ln  ({m_\Pl}/{m_e})  =3/8$ to lowest order, and attempted to improve it by a more complete description of the transition occuring around the electron Compton scale.

Then we have suggested the existence of a similar relation for the lightest neutrino, reading $\alpha  \ln  ({m_\Pl}/{m_\nu} ) =1/2$. From this relation, we obtain a theoretical prediction for the lightest neutrino mass $m_{\nu}=m_\Pl \: \exp (-\frac{1}{2}\alpha^{-1})=0.0214$ eV. This corresponds to an inverse ratio $2.39 \times 10^7$ with the electron mass, i.e. to a Compton length $\lambda_\nu=\hbar/m_\nu c=9.22\: \mu$m, of the order of size of a living cell.

 The two other neutrino masses can be derived from this mass and from experimental results of neutrino oscillations. We find $m_{\nu2}=0.0231 \:{\rm eV}$ and $m_{\nu3}=0.0552 \:{\rm eV}$, yielding a total mass of $0.1$ eV, compatible with the current experimental limits.  

It will be possible to put this expectation to the test by future neutrino experiments allowing direct mass measurements. The present upper limit by KATRIN, $m_\nu<1.1$ eV  \cite{Aker2019}, is still $\approx 50$ times the mass predicted in the present work. If no neutrino mass signal is found, the KATRIN sensitivity after 3 years of measurements will be $m_\nu<0.2$ eV ($90\%$ CL.), which is closer to our theoretical prediction but still larger. 

%%%%%%%%%%%%

%%%%%%%%%%%
%%%%%%
\end{document}